\documentclass{article}

\usepackage{arxiv}

\usepackage{bm}
\usepackage{amssymb, amsmath} 
\usepackage{multirow} 
\usepackage{hyperref} 
\usepackage{graphicx} 

\usepackage[utf8]{inputenc} 
\usepackage[T1]{fontenc}    
\usepackage{hyperref}       
\usepackage{url}            
\usepackage{booktabs}       
\usepackage{amsfonts}       
\usepackage{nicefrac}       
\usepackage{microtype}      
\usepackage{cleveref}       
\usepackage{lipsum}         
\usepackage{graphicx}
\usepackage{natbib}
\usepackage{doi}

\title{Defiltering turbulent flow fields for Lagrangian particle tracking using machine learning techniques\thanks{
This article has been accepted by \textit{Physics of Fluids}. After it is published, it will be found at \url{https://pubs.aip.org/aip/pof}.}}

\newif\ifuniqueAffiliation
\uniqueAffiliationtrue

\ifuniqueAffiliation 
\author{
	\textbf{Tomoya Oura}\\
	Department of Mechanical Engineering\\
	Keio University\\
	Yokohama, 223-8522, Japan\\
	\texttt{oura.tomoya@keio.jp}\\
	\and
	\textbf{Koji Fukagata}\\
	Department of Mechanical Engineering\\
	Keio University\\
	Yokohama, 223-8522, Japan
}

\renewcommand{\shorttitle}{}

\hypersetup{
pdftitle={Defiltering turbulent flow fields for Lagrangian particle tracking using machine learning technique},
pdfsubject={physics},
pdfauthor={Tomoya Oura, Koji Fukagata},
}

\begin{document}
\maketitle

\begin{abstract}
	We propose a defiltering method of turbulent flow fields for Lagrangian particle tracking using machine learning techniques. Numerical simulation of Lagrangian particle tracking is commonly used in various fields. In general, practical applications require an affordable grid size due to the limitation of computational resources; for instance, a large-eddy simulation reduces the number of grid points with a filtering operator. However, low resolution flow fields usually underestimate the fluctuations of particle velocity. We thus present a novel approach to defilter the fluid velocity to improve the particle motion in coarse-grid (i.e., filtered) fields. The proposed method, which is based on the machine learning techniques, intends to reconstruct the fluid velocity at a particle location. We assess this method in {\it a priori} manner using a turbulent channel flow at the friction Reynolds number ${\rm Re}_\tau=$ $180$. The investigation is conducted for the filter size, $n_{\rm filter}$, of $4$, $8$, and $16$. In the case of $n_{\rm filter} = 4$, the proposed method can perfectly reconstruct the fluid velocity fluctuations. The results of $n_{\rm filter} = 8$ and $16$ also exhibit substantial improvements in the fluctuation statistics although with some underestimations. Subsequently, the particle motion computed using the present method is analyzed. The trajectories, the velocity fluctuations, and the deposition velocity of particles are reconstructed accurately. Moreover, the generalizability of the present method is also demonstrated using the fields whose computational domain is larger than that used for the training. The present findings suggest that machine learning-based velocity reconstruction will enable us precise particle tracking in coarse-grid flow fields. 
\end{abstract}


\section{Introduction}
\label{sec:introduction}

Transport of inertial particles in turbulent flows is a crucial topic in various fields, such as industry~\citep{wang_numerical_2006, de_souza_large_2012}, pharmacy~\citep{kleinstreuer_targeted_2008, kleinstreuer_airflow_2010}, and epidemiology~\citep{yang_effects_2018, saw_modeling_2021, rencken_patterns_2021}. To understand and analyze it, a lot of experiments and observations have been conducted in the past. However, motion of each particle can hardly be observed in many practical applications due to the temporal and/or spatial resolution of the measuring instruments. In such cases, a numerical simulation that can track the individual particles serves as a powerful tool. The method to track the motion of discrete particles in continuous fields is known as Lagrangian particle tracking (LPT).

The computational domain should usually be taken large enough in numerical simulations of practial flows. For example, in a mixing tank simulation, the entire region of the tank needs to be considered. These cases usually require compuations with an affordable grid size due to the limitation of computational resources, such as the large-eddy simulation (LES) considering the filtered velocity fields. However, it is well known that the coarse-grid simulation underestimates the velocity fluctuations, the dispersion, and the accumulation of inertial particles~\citep{armenio_effect_1999, kuerten_can_2005, shotorban_improvement_2007}. The underestimation of the fluid velocity fluctuations directly affects the particle velocity, and causes the inaccurate motion of particles.

To overcome this problem, several methods to improve particle motion in coarse-grid fields have been proposed. These methods are broadly classified into two types~\citep{kuerten_point-particle_2016, marchioli_large-eddy_2017}: the defiltering approach and the stochastic approach. The defiltering approach is an idea to reconstruct the correct velocity from the filtered (i.e., the coarse-grid) fields. One example is the fractal interpolation (FI)~\citep{scotti_meneveau_1999}, which is a simple way to interpolate the fluid velocity at a given point. However, because FI assumes the parameters obtained in a homogeneous flow, this method was reported insufficient to reconstruct the fluid velocity of the turbulent channel flow, particularly near the wall~\citep{marchioli_appraisal_2008}. Another option of the defiltering approach is the approximate deconvolution method (ADM)~\citep{stolz_adams_1999}, which approximates the inverse filtering operator, $Q_\alpha$, using the van Cittert series as
\begin{eqnarray}
    Q_\alpha = \sum_{k = 0}^{\alpha} (1 - G)^{k}
    \label{eq:ADM}
    ,
\end{eqnarray}
where $G$ is a filtering operator.
It has been applied to the turbulent channel flow~\citep{kuerten_can_2005, kuerten_subgrid_2006, marchioli_appraisal_2008} and the homogeneous shear turbulence~\citep{shotorban_modeling_2005, shotorban_improvement_2007}. This method showed remarkable success for the moderate filter size, but it did not work well for very coarse fields~\citep{marchioli_appraisal_2008}. Moreover, this method restricts that the filter kernel must be known. The second approach, i.e., the stochastic approach, was proposed by Pozorski and Minier~\citep{pozorski_probability_1999}, and it has been applied to the flow around a bluff-body~\citep{minier_pdf_2004} and a turbulent pipe flow~\citep{chibbaro_langevin_2008}. Because this is a statistical approach based on the Langevin equation, it succeeded in obtaining the one-point statistical moments, whereas it could not reconstruct the motion of individual particles~\citep{marchioli_large-eddy_2017}. As mentioned above, much effort has been made to improve the particle motion in the coarse-grid simulation, but issues still remain for practical computations on very coarse grids.

Nowadays, machine learning (ML) techniques have been widely used in fluid mechanics~\citep{brunton_2020,vinuesa_brunton_2022,vinuesa_2023,fukagata_2023}. Because ML contains nonlinear functions, it can well adjust to nonlinear behavior of fluid flows. One of the noteworthy successes is the super-resolution reconstruction of turbulent flows. Fukami et al. showed that the super-resolution technique with ML can be applied to the two-dimensional~\citep{fukami_super-resolution_2019} and the three-dimensional turbulent flows~\citep{fukami_machine-learning-based_2021} to reconstruct the high-resolved fields from the coarse-grid fields. In terms of LES, subgrid-scale modeling with ML techniques has also been investigated~\citep{gamahara_searching_2017, wang_investigations_2018, beck_deep_2019}. The models in these studies have attempted to estimate the subgrid-scale stress tensor from filtered fields using ML instead of the conventional method such as the Smagorinsky model~\citep{smagorinsky_general_1963}. These studies are conducted in a supervised manner, in other words, the training processes of the model require high-resolved reference data. 

Motivated by the studies above, here we propose an ML-based defiltering method of turbulent flow fields for the Lagrangian particle tracking (LPT). We demonstrate our method, which reconstructs the fluid velocity at a given point in the filtered field using local information around the point, by conducting {\it a priori} test. We also conduct LPT using the defiltered fluid velocity and investigate the recovering accuracy. Finally, to indicate the generalizability of the present method, we apply the model which is trained in the minimal computational domain to the field in a larger domain.

\section{Methods}
\label{sec:methods}

\subsection{Turbulent channel flow}
\label{sec:turbulent_channel_flow}

We consider a three-dimensional velocity field of a turbulent channel flow at ${\rm Re}_{\tau}=180$, where ${\rm Re}_{\tau} = u_{\tau} h / \nu$ denotes the friction Reynolds number based on the friction velocity, $u_{\tau}$, the channel half-width, $h$, and the kinematic viscosity, $\nu$. The governing equations are the incompressible continuity equation and the Navier-Stokes equation, 
\begin{eqnarray}
    {\bm \nabla} \cdot {\bm u}_f &=& 0 
    \label{eq:continuity_eq} \\
    \frac{\partial {\bm u}_f}{\partial t} &=& -{\bm \nabla} \cdot \left( {\bm u}_f {\bm u}_f \right) - {\bm \nabla} p + \frac{1}{{\rm Re}_{\tau}} {\bm \nabla}^2 {\bm u}_f
    \label{eq:NS_eq}
    ,
\end{eqnarray}
where ${\bm u}_f$, $p$, and $t$ denote the fluid velocity, pressure, and time, respectively. The fluid velocity vector, ${\bm u}_f$, is composed of $u_f$, $v_f$, and $w_f$, i.e., the streamwise ($x$), wall-normal ($y$), and spanwise ($z$) components. Hereafter, the quantities in wall units (the non-dimensionalization with $u_{\tau}$ and $\nu$) are represented by a superscirpt of $+$.

\begin{table}[t]
    \caption{Computational domains of DNS.}
    \label{table:computational_domains}
    \hspace{100truemm}
    \centering
    {\tabcolsep=2mm
        \begin{tabular}{cccccc}
            \hline \hline
            Domain & $(L_x, L_y, L_z)$ & Filter & $(N_x, N_y, N_z)$ & $\Delta x^+$ & $\Delta z^+$ \\
            \hline
            \multirow{4}{*}{Minimal} & \multirow{4}{*}{$(\pi h, 2h, \pi h /2)$} & Original & $(64, 96, 64)$ & $8.84$ & $4.42$ \\ 
             & & $n_{\rm filter} = 4$ & $(16, 24, 16)$ & $35.3$ & $17.7$ \\
             & & $n_{\rm filter} = 8$ & $(8, 12, 8)$ & $70.7$ & $35.3$ \\
             & & $n_{\rm filter} = 16$ & $(4, 6, 4)$ & $141$ & $70.7$ \\
            \hline
            \multirow{2}{*}{Larger} & \multirow{2}{*}{$(4\pi h, 2h, 2\pi h)$} & Original & $(256, 96, 256)$ & $8.84$ & $4.42$ \\ 
            & & $n_{\rm filter} = 4$ & $(64, 24, 64)$ & $35.3$ & $17.7$ \\
            \hline \hline
        \end{tabular}
    }
\end{table}

First, direct numerical simulation (DNS)~\citep{fukagata_theoretical_2006}, which solves Eqs.~(\ref{eq:continuity_eq})--(\ref{eq:NS_eq}), is performed to obtain the correct fluid velocity field. As summarized in Table~\ref{table:computational_domains}, two different sizes of the computational domain are considered in the present study: the ``minimal'' domain, $(L_x, L_y, L_z) = (\pi h, 2h, \pi h / 2)$, which will be used for the training of the ML model, and the ``larger'' domain, $(L_x, L_y, L_z) = (4\pi h, 2h, 2\pi h)$, where the generalizability of the developed ML model will be examined. The number of grid points is $(N_x, N_y, N_z) = (64, 96, 64)$ in the minimal domain, and $(N_x, N_y, N_z) = (256, 96, 256)$ in the larger domain. The computational grids are taken uniformly in $x$ and $z$ directions, and ununiformly in $y$ direction. The minimal domain was justified in several studies~\citep{jimenez_minimal_1991, nakamura_convolutional_2021} as maintaining coherent structures of turbulence. The time-step size of the fluid flow calculation, $\Delta t_f^+$, is set to $0.045$ in both cases. 

The coarse-grid fields are obtained by filtering the DNS data offline. Namely, the fluid velocity fields are solved in advance with the original resolution, and then, a spatial filter is applied to the field at each time step. The filter size, $n_{\rm filter}$, is set to $4$, $8$, $16$ for the minimal domain, and $4$ for the larger domain. As the filtering operator, a box filter (i.e., the average pooling with the kernel size of $n_{\rm filter} \times n_{\rm filter} \times n_{\rm filter}$ and the stride size of $n_{\rm filter}$) is used.

\subsection{Machine learning based defiltering method}
\label{sec:ML_based_defiltering_method}

\subsubsection{Model structure}
\label{sec:model_structure}

\begin{figure}[b]
    \centering
    \includegraphics[width=130mm]{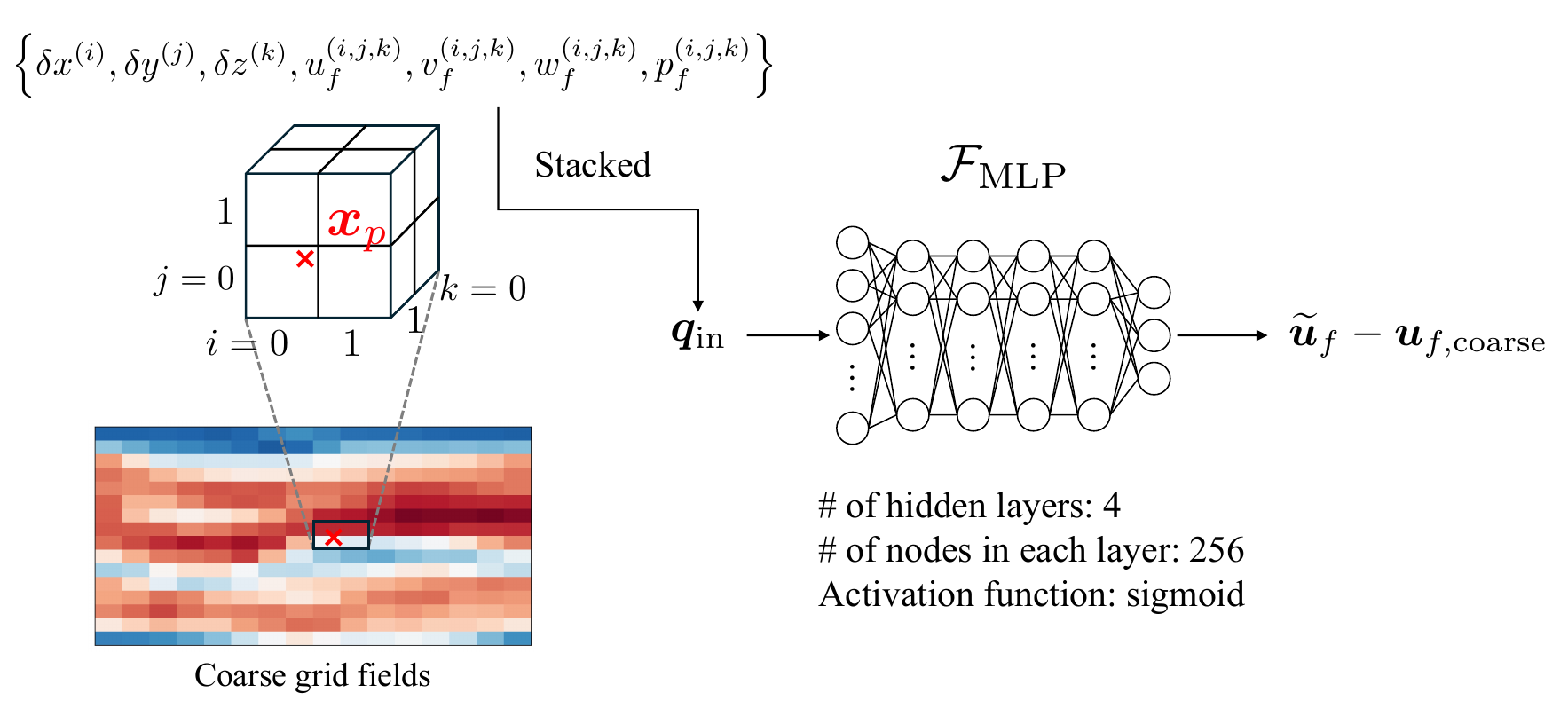}
    \caption{The schematic drawing of the present method.}
    \label{fig:schematic_view}
\end{figure}

We use a multi-layer perceptron (MLP)~\citep{rumelhart1986learning} as the ML model to reconstruct the fluid velocity at the particle location in the coarse-grid fields. The operation in the MLP is represented as
\begin{eqnarray}
    s_i^{(l)} = \phi \left( \sum_j w_{ij}^{(l)} s^{(l-1)} + b^{(l)}_j \right)
\label{eq:general_MLP}
,
\end{eqnarray}
where $s_i^{(l)}$ denotes the node of a network in the $l$-th layer, $w_{ij}^{(l)}$ and $b_j^{(l)}$ are weights and biases, and $\phi$ is an activation function. The weights and biases are optimized through the back-propagation to fit the given data. The present model contains four hidden layers with 256 nodes each. As the activation function, the sigmoid function, $\phi(\xi) = 1 / \left(1 + e^{-\xi} \right)$, is used for the hidden layers, whereas the linear function is used for the output layer. The comparison regarding the model sturctures and activation functions will be discussed in Section~\ref{sec:model_structure_assessment}.

The input to the present ML model is the local information around the particle location. We train the model to output the difference between the interpolated velocity in the coarse-grid fields and that in the fine-grid fields. The model can be represented as
\begin{eqnarray}
    {\bm u}_{f, {\rm fine}} \left( {\bm x}_p\right) - {\bm u}_{f, {\rm coarse}} \left( {\bm x}_p \right) = \mathcal{F}_{\rm MLP} \left( {\bm q}_{\rm in} \right)
\label{eq:output_of_MLP}
.
\end{eqnarray}
On the left-hand side, ${\bm u}_{f, {\rm fine}} \left( {\bm x}_p \right)$ and ${\bm u}_{f, {\rm coarse}} \left( {\bm x}_p \right)$ are the fluid velocities interpolated onto the particle location, ${\bm x}_p$, from the fine and coarse grids, respectively. For both grids, the interpotaion is done by using the fourth-order Lagrangian interpolation, which has commonly be used as a high accuracy method in LPT simulations since the early studies\citep{wang_squires_1996}. On the right-hand side, $\mathcal{F}_{\rm MLP}$ denotes the MLP, and ${\bm q}_{\rm in}$ denotes the input variables from the coarse-grid fields, which are discussed in detail below. Once the weights and biases of $\mathcal{F}_{\rm MLP}$ are optimized using training data (i.e., the ground truth data), we can obtain the defiltered fluid velocity, $\widetilde{\bm{u}}_{f}$, at a given location, $\bm{x}_p$, as
\begin{eqnarray}
    \widetilde{\bm{u}}_f \left(\bm{x}_p \right) =  {\bm u}_{f, {\rm coarse}} \left( {\bm x}_p \right) + \mathcal{F}_{\rm MLP} \left( {\bm q}_{\rm in} \right)
\label{eq:defiltered_velocity_is_equal_to}
.
\end{eqnarray}

For the input quantities, ${\bm q}_{\rm in}$, the grid points around the particle location in the coarse-grid fields are considered. Figure~\ref{fig:schematic_view} shows the schematic of the present method. Around the particle location, ${\bm x}_p = (x_p, y_p, z_p)$, we select the two grid points in each direction, labeled as $(i,j,k)$, for $i,j,k = 0, 1$. The MLP obtain the relative distance, $( \delta x^{(i)}, \delta y^{(j)}, \delta z^{(k)} )$, the fluid velocity, ${\bm u}_f^{(i,j,k)}$, and the fluid pressure, $p_f^{(i,j,k)}$, at each grid point. Here, the relative distance means the distance between the grid point and the particle location. For example, the relative distance in $x$-direction is defined as
\begin{eqnarray}
    \delta x_f^{(i)} \equiv x^{(i)}_f - x_p
    \label{eq:relative_distance}
\end{eqnarray}
for $i = 0, 1$. All variables are stacked to form an input vector, ${\bm q}_{\rm in} \in \mathbb{R}^{38}$--- 8 neighboring points times 4 quantities ($u_f$, $v_f$, $w_f$, $p_f$), plus 2 grid points for 3 directions for the relative distance ---, and given to the input layer of the function, $\mathcal{F}_{\rm MLP}$.

\subsubsection{Training strategy}
\label{sec:training_strategy}
The DNS of turbulent channel flow, described in Section~\ref{sec:turbulent_channel_flow}, creates the training and validation data for the present ML model. Every $\Delta t^+ = 1.8$, fluid velocities at randomly selected points, $\left\{ {\bm x}_p \right\}$, are interpolated from the DNS grids to obtain ${\bm u}_{f, {\rm fine}} \left( {\bm x}_p \right)$. For the reason discussed in Section~\ref{sec:model_structure_assessment} later, 6144 points are taken in the bulk region ($10 \leq y_p^+ \leq 350$) and 1000 points are taken in the near-wall region ($y_p^+ < 10$ and $y_p^+ > 350$). To obtain ${\bm u}_{f,{\rm coarse}}$ and $p_{f,{\rm coarse}}$ at the same $\left\{ {\bm x}_p \right\}$, the fluid velocities and pressure are also interpolated from the filtered field. As mentioned above, the ground truth is given by ${\bm u}_{f, {\rm fine}} - {\bm u}_{f, {\rm coarse}}$, while the input data $\left\{ {\bm q}_{\rm in} \right\}$ consist of ${\bm u}_{f, {\rm coarse}}$ and $p_{f, {\rm coarse}}$ around $\left\{ {\bm x}_p \right\}$ as well as the relative distances. The DNS is conducted until $t^+ = 3600$. 

Model training is performed using the Adam optimizer~\citep{kingma2017adam}, and the loss is evaluated with the mean squared error (MSE). During training, dataset is divided into mini-batches whose batch size is set to $4096$. The number of epochs is determined to be 600 based on the loss convergence. To assess the variation of the prediction accuracy, the five-fold cross-validation~\citep{fukami_assessment_2020} is conducted.

\subsection{Lagrangian particle tracking (LPT)}
\label{sec:Lagrangian_particle_tracking}

The particles are described as discrete points in the continuous fluid fields. 
By assuming a dilute dispersion of monodispersed spherical particles, the coordinate, ${\bm x}_{p, i}$, and velocity, ${\bm u}_{p, i}$, of the $i$-th particle is governed by equations of motion, i.e.,
\begin{eqnarray}
    \frac{d {\bm x}_{p, i}}{dt} &=& {\bm u_{p, i}}
    \label{eq:eom_of_particle_x}
    , \\
    \frac{d {\bm u}_{p, i}}{dt} &=& \frac{1}{\tau_p}\left({\bm u}_f \left({\bm x}_{p, i} \right) - {\bm u}_{p, i}\right)
    \label{eq:eom_of_particle_u}
    .
\end{eqnarray}
Hereafter, the subscript, $i$, will be omitted for a notational simplicity. The Stokes relaxation time, $\tau_p$, is defined as~\citep{wood_mass-transfer_1981}
\begin{eqnarray}
    \tau_p = \frac{d^2 S}{18 \nu}
    \label{eq:tau_p}
    ,
\end{eqnarray}
where $d$ is the particle diameter, and $S \equiv \rho_p / \rho_f$ is the density ratio of particles to fluid. For simplicity, only the drag force is considered in this study. Note that the lift force may affect the particle deposition~\citep{marchioli_influence_2007} although its contribution is usually smaller than the drag force. Since the present method is applicable as it is even when the lift force term is present, we simply ignore it in the present study. Other typical forces, such as added mass, Basset force, pressure gradient, and Fax\'en correction, are known to be negligible for $\sim 10$ $\mu$m solid particles in air as we consider as an example in the present study~\citep{fan_sublayer_1993}. When LPT is conducted without using the present defiltering method, the fluid velocity at the particle location, ${\bm u}_f \left({\bm x}_p \right)$, is calculated with the fourth-order Lagrangian interpolation, as has been a common practice in LPT simulations coupled with LES\citep{wang_squires_1996}. With the proposed defiltering method, in contrast, the defiltered fluid velocity, $\widetilde{{\bm u}}_f \left({\bm x}_{p} \right)$ obtained by Eq. (\ref{eq:defiltered_velocity_is_equal_to}), is used to substitute ${\bm u}_f \left({\bm x}_p \right)$ in Eq.~(\ref{eq:eom_of_particle_u}); namely,
\begin{eqnarray}
    \frac{d {\bm u}_{p}}{dt} = 
    \frac{1}{\tau_p}
    \left(
    \widetilde{{\bm u}}_f \left({\bm x}_{p} \right) - {\bm u}_{p}
    \right) 
    .
    \label{eq:EoM_with_defiltered_velocity}
\end{eqnarray}
For the initial condition, $3600$ particles are randomly introduced in the computational domain. The initial velocity of particles is set equal to the fluid velocity at the location. Trajectories of particles are calculated with Eq.~(\ref{eq:eom_of_particle_x}) and Eq.~(\ref{eq:eom_of_particle_u}) or (\ref{eq:EoM_with_defiltered_velocity}). Time integration is performed using the Crank-Nicolson method for the coordinates and the third-order Adams-Bashforth method for the velocity. Note that, in the first two time-steps, the Euler implicit method is used for integrating the velocity. 

In this study, moderately inertial particles are considered. The non-dimensionalized Stokes relaxation time, which is equal to the Stokes number, ${\rm St}$, is set to $\tau_p^+ = {\rm St} = 10.0$. Note that particles having this order of relaxation time ($\tau_p^+ \sim 10.0$) are of engineering importance, as they result in the locally maximum deposition rate in wall-bounded turbulent flows~\citep{wood_mass-transfer_1981}. For simplicity, the interactions of particles to fluid as well as those between particles to particles are neglected (i.e., one-way coupling). Note that the volume fraction of particles is approximately $8 \times 10^{-7}$, which can justify the one-way coupling assumption. The time step size for LPT is set to $\Delta t^+_p = 0.18$. We confirmed that no differences were observed between the results of $\Delta t^+_p = 0.045$ and $0.18$, although the detailed results are not presented here for brevity. The total tracking time is $t^+ = 4815$. The periodic boundary condition is applied to the particles that reach the boundaries in the $x$ and $z$ directions. In the $y$ direction, the particles are judged to deposit to the wall as soon as they reach the boundaries and removed from the computational domain. 

\section{Results and discussion}
\label{sec:results_and_discussion} 

In this section, we assess the present method by comparing the defiltered results with the fine-grid ones. The investigations in Sections~\ref{sec:model_structure_assessment}--\ref{sec:characteristics_of_tracked_particles} are carried out in the computational domain whose size is same as that used for training the ML model. In Section~\ref{sec:fluid_velocity_statistics_in_larger_domain}, the generalizability of the present ML model is assessed using the domain which is larger than that used for training. It should be emphasized that the fluid fields for validation are taken from a time interval sufficiently far apart, and they are completely uncorrelated with those used for training the ML model. 

\subsection{Assessment of the model structure and the training data amount}
\label{sec:model_structure_assessment} 

\begin{figure}[t]
    \centering
    \includegraphics[width=100mm]{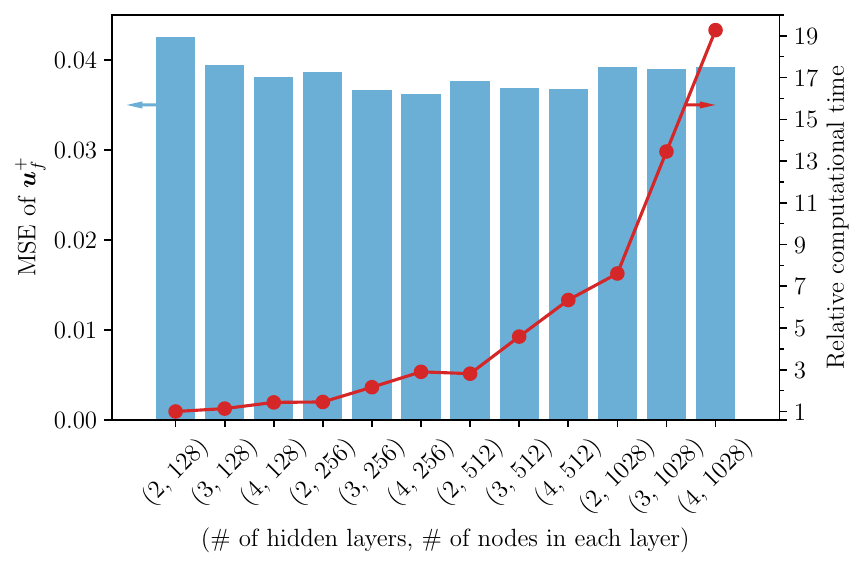}
    \caption{The dependence of MSE (bars on the left axis) and computational time (line plots on the right axis) on the model structure. The horizontal axis represents the number of hidden layers and the number of nodes. For instance, $(2, 128)$ denotes the case of 2 hidden layers with 128 nodes in each layer. The computational time is normalized by the value in the case of $(2, 128)$.}
        \label{fig:network_structures}
\end{figure}

\begin{table}[b]
    \caption{The MSE of fluid velocity among five activation functions.}
    \label{table:comparison_of_activation_functions}
    \hspace{100truemm}
    \centering
    {\tabcolsep=5mm
        \begin{tabular}{ccccc}
            \hline \hline
            Activation function & Formula & MSE \\
            \hline
            Linear & $\phi(\xi) = \xi$ & $0.0697$ \\
            ReLU & $\phi(\xi) = \max (0, \xi)$ & $0.0384$ \\
            Sigmoid & $\phi(\xi) = 1 / (1 + e^{-\xi})$ & $0.0362$ \\
            Softplus & $\phi(\xi) = \ln ( e^{\xi} + 1)$ & $0.0366$ \\
            tanh & $\phi(\xi) = (e^{\xi} - e^{-\xi}) / (e^{\xi} + e^{-\xi})$ & $0.0383$ \\
            \hline \hline
        \end{tabular}
    }
\end{table}

First, we assess the ML-model structure of the present method, namely, the number of hidden layers, the number of nodes, and the activation function. To compare the model structures, the defiltering accuracy is evaluated using the MSE between the original fluid velocity, ${\bm u}_{f, {\rm fine}}$, and the defiltered fluid velocity, $\widetilde{{\bm u}}_f$. For this assessment, the number of training data points, $\left\{ {\bm x}_p \right\}$, is set to 6144 in the entire domain, and the snapshots are taken every $\Delta t^+ = 1.8$ for $t^+ = 3600$. To save computational time, no cross-validation is conducted, and the early-stopping criteria~\citep{prechelt_automatic_1998} is used to avoid overfitting.

Figure~\ref{fig:network_structures} shows the defiltering accuracy and the training time depending on the number of hidden layers and the number of nodes in each layer. Here, the sigmoid function is used as the activation function. Although the MSE decreases as the numbers of hidden layers and nodes increase for smaller networks, the reduction rate is saturated around four hidden layers and 256 nodes and even deteriorates for larger networks. In contrast, it takes longer time to train larger networks. Judging from the balance between the accuracy and the training time, the model containing four hidden layers with 256 nodes each is used in this study.

Next, we assess five different activation functions, i.e., the linear function, the Rectified Linear Unit (ReLU), the sigmoid function, the softplus function, and the hyperbolic tangent function, which are applied to all the hidden layers except for the last layer where the linear activation is used. Table~\ref{table:comparison_of_activation_functions} shows the comparison of activation functions. The linear activation shows the lowest accuracy, which indicates that the nonlinear activation function plays a key role in the present method. Among the nonlinear activation functions, the sigmoid function shows the best accuracy for the present problem. Accordingly, we select the sigmoid function as the activation function in this study.

\begin{figure}[b]
    \centering
    \includegraphics[width=100mm]{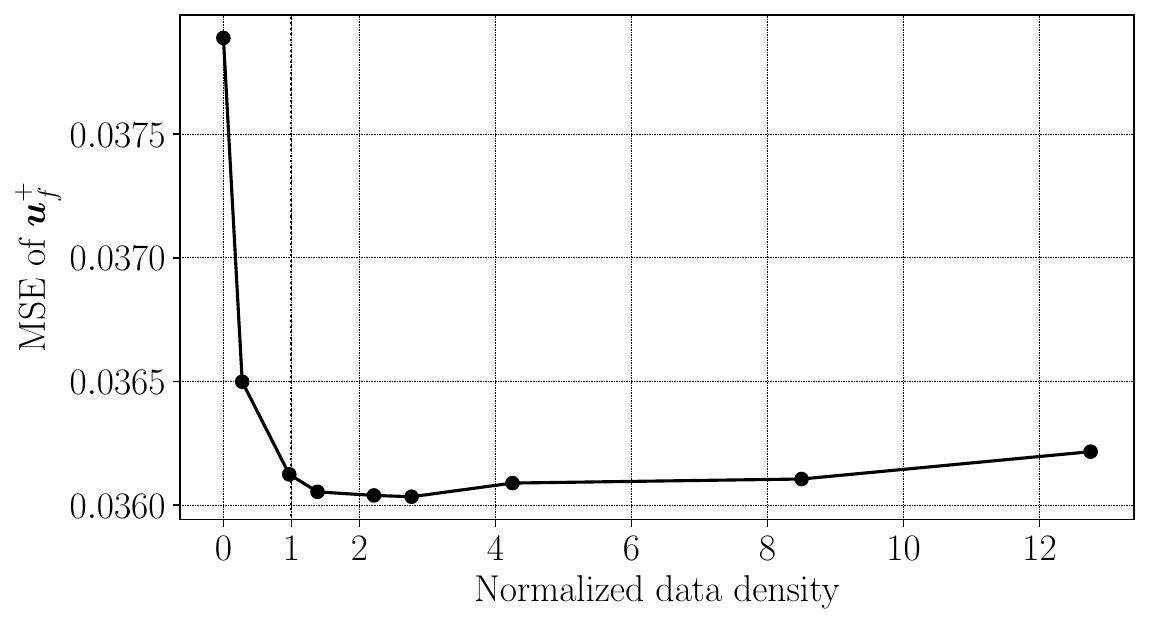}
    \caption{Effect of the training data density in the near-wall region.
          The density in the near-wall region is normalized by 
          that in the bulk region.}
    \label{fig:training_data_density}
\end{figure}

Subsequently, the assessment is conducted on the amonut of training data. Considering the substantial difference in the turbulence dynamics, we divide the channel domain into two parts, the near-wall region ($y^+ < 10$ and $y^+ > 350$) and the bulk region ($10 \leq y^+ \leq 350$). We evaluate the contribution of the amount of training data in the near-wall region while that in the bulk region is fixed. Figure~\ref{fig:training_data_density} shows the effect of the amount of training data in the near-wall region on the defiltering accuracy. The horizontal axis is the data density in the near-wall region normalized by that in the bulk region --- unity means that all training data points are distributed uniformly in the entire domain. The result implies that enhancing the training data density in the near-wall region improves the accuracy, whereas the accuracy gets worse with too much data in the near-wall region. The model performs best when the amount of data in the near-wall region is about 2.8 times more than that in the bulk region, although the dependency around this point is mild. From this result, we use 6144 points in the bulk region and 1000 points in the near-wall region, which is 2.8 times more than that in the bulk region based on the density.

\subsection{Statistics of fluid velocity in the minimal domain}
\label{sec:fluid_velocity_statistics_in_minimal_domain}

As explained above, the training of the ML model is done using the data obtained from the DNS conducted in the minimal domain. Therefore, in this section, the fluid velocity statistics in the minimal domain are presented as the baseline assessment.

\begin{figure}[t]
    \centering
    \includegraphics[width=130mm]{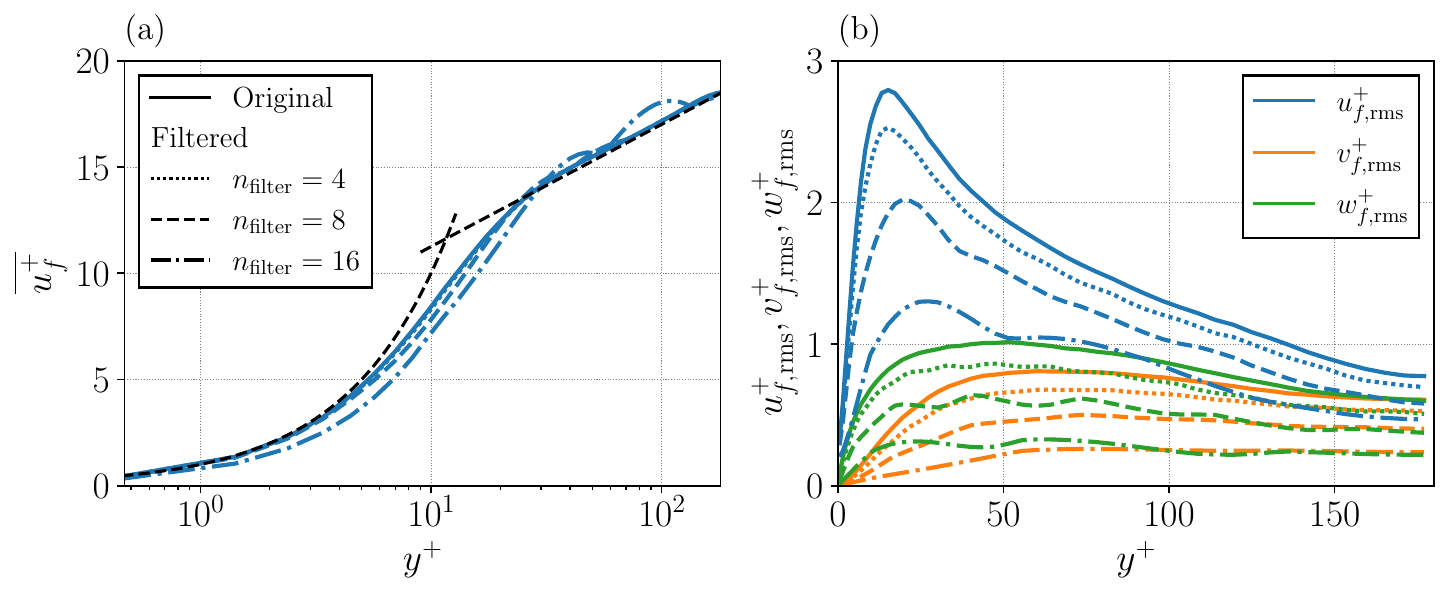}
    \caption{Statistics of the original and filtered velocities at particle locations: (a) The mean streamwise velocity; (b) RMS of fluid velocity fluctuations. The black dashed lines in the figure (a) represent $\overline{u^+_f} = y^+$ and $\overline{u^+_f} = \ln \ y^+ / 0.4 + 5.5$. These are calculated in a minimal domain.}
    \label{fig:stat_of_fluid_vel_original_and_filtered}
\end{figure}

Figure~\ref{fig:stat_of_fluid_vel_original_and_filtered} shows the mean and the root-mean-square (RMS) of the fluid velocity fluctuations in the fine-grid (i.e., the original DNS) and the coarse-grid (i.e., the filtered DNS) fields. Figure~\ref{fig:stat_of_fluid_vel_original_and_filtered} (a), showing the mean streamwise velocity profiles, demonstrates the followings: (1) the profiles reasonably follow the linear law ($\overline{u^+_f} = y^+$) and the log law ($\overline{u^+_f} = {\rm ln} \ y^+ / 0.4 + 5.5$) even if the minimal domain is adopted; (2) no difference between the fine and the coarse-grid fields is observed in $n_{\rm filter} = 4$; and (3) the results of $n_{\rm filter} = 8$ and $16$ underestimate the mean velocity profile. In contrast, Fig.~\ref{fig:stat_of_fluid_vel_original_and_filtered} (b) indicates that the coarse-grid simulation underestimates the RMS fluctuations of each velocity component and the discrepancies become larger as the filter size increases.

\begin{figure}[t]
    \centering
    \includegraphics[width=130mm]{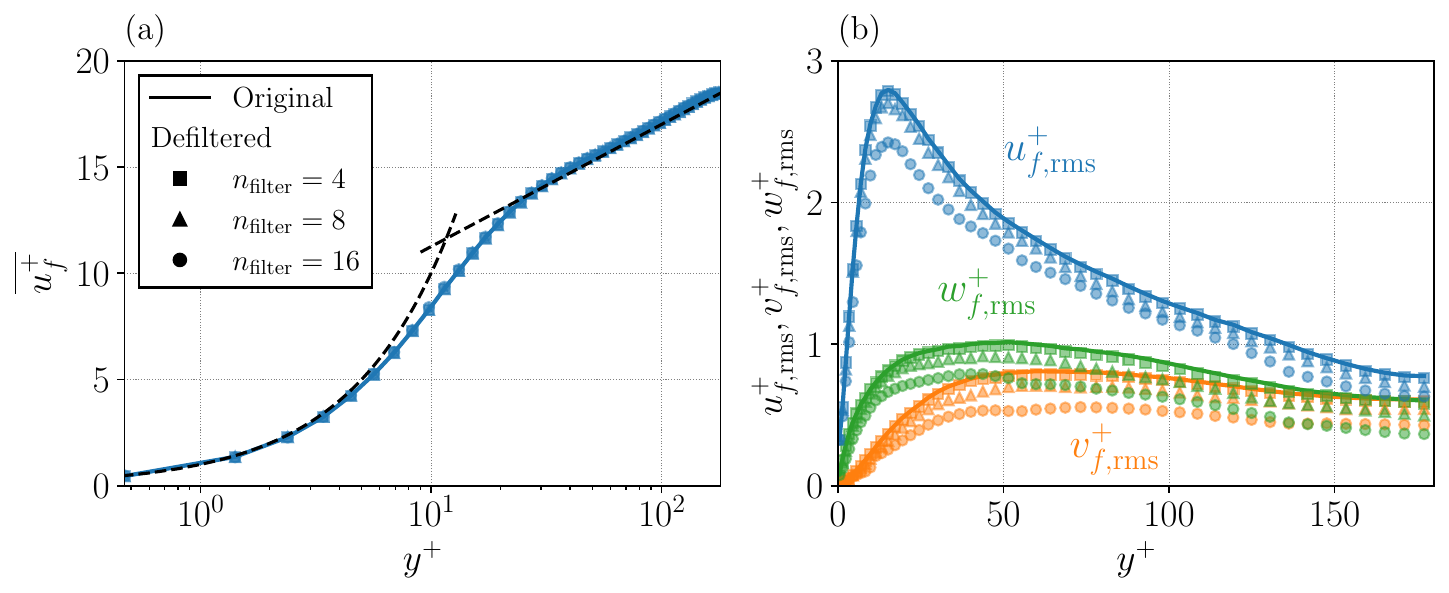}
    \caption{Statistics of the defiltered velocities at particle locations: (a) The mean streamwise velocity; (b) RMS of fluid velocity fluctuations. These are calculated in a minimal domain.}
    \label{fig:stat_of_fluid_vel_original_and_defiltered}
\end{figure}

We subsequently assess the statistics of the fluid velocity defiltered with the present method. The mean velocity and the RMS profiles of the defiltered fluid velocity compared to the original one are presented in Fig.~\ref{fig:stat_of_fluid_vel_original_and_defiltered}. Figure~\ref{fig:stat_of_fluid_vel_original_and_defiltered} (a) demonstrates that the present models of all filter sizes can perfectly reconstruct the mean velocity profile. In Fig.~\ref{fig:stat_of_fluid_vel_original_and_defiltered} (b) the RMS profiles of the reconstructed velocity show great agreement with the original profiles when $n_{\rm filter} = 4$. The result of $n_{\rm filter} = 8$ is only slightly underestimated. The reconstruction accuracy of $n_{\rm filter} = 16$ is insufficient, but substantially improved as compared to the original profiles presented in Fig.~\ref{fig:stat_of_fluid_vel_original_and_filtered} (b). Hereafter, we present the results of in the case of $n_{\rm filter} = 4$ only.

\subsection{Characteristics of the tracked particles}
\label{sec:characteristics_of_tracked_particles}

\begin{figure}[t]
    \centering
    \includegraphics[width=90mm]{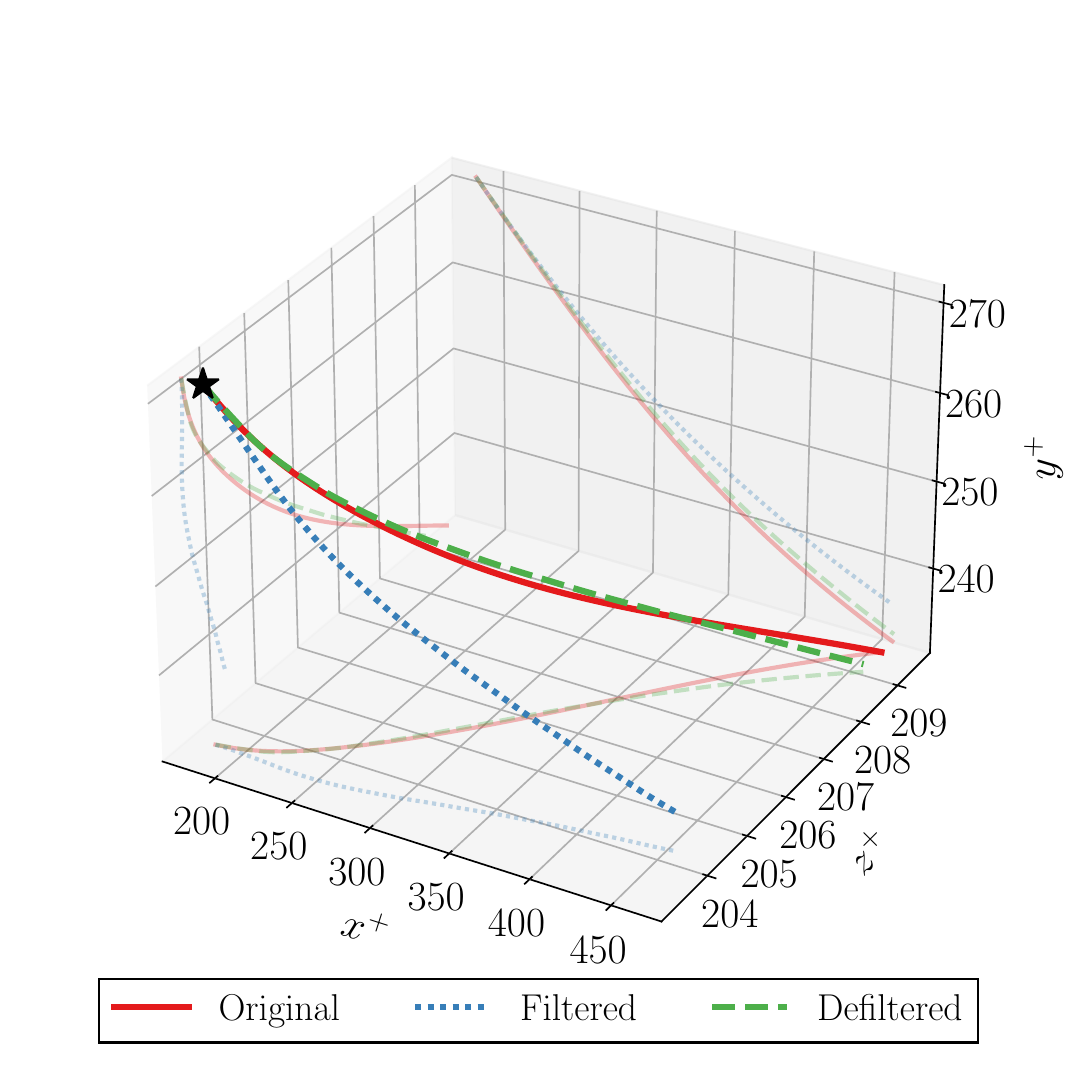}
    \caption{The particle trajectories of one particle for $t^+ = 0$--$18$. The star point is the initial location of the particle. The two-dimensional projections are also plotted on each plane.}
    \label{fig:particle_trajectories}
\end{figure}

Particle motions in the fine-grid, coarse-grid (i.e., filtered), and defiltered fields are investigated based on the LPT conducted in the fluid fields of the minimal domain. It should be emphasized that the ``coarse-grid'' means the result of the fourth-order Lagrangian interpolation from the coarse grid, i.e., the method commonly used in LES of gas-particle flows~\citep{wang_squires_1996}. In some investigations, the results of two other methods, such as the ``linear'' method and the ADM, are also shown for comparison. The ``linear'' results are obtained using the model with the same network structure as the present ML model but with the linear activation (i.e., instead of the sigmoid function). For the ADM, which is represented as Eq.~(\ref{eq:ADM}), $\alpha$ is set to $5$ and the Gaussian filter whose filter size is twice as large as $n_{\rm filter}$ is used as the filtering operator, $G$, following the previous study~\citep{kuerten_subgrid_2006, shotorban_improvement_2007}.

First, the three-dimensional trajectories of a particle are exemplified in Fig.~\ref{fig:particle_trajectories}. These trajectories are plotted for $t^+ = 0$--$18$, which corresponds to the initial 100 time-steps. The two-dimensional projections are also plotted on each wall. For better readability, the trajectories across the periodic boundaries are drawn as connected, and the scales of each direction are not equal. This result shows that the present method can reconstruct the particle motion accurately thanks to the defiltered fluid velocity.

\begin{figure}[t]
    \centering
    \includegraphics[width=80mm]{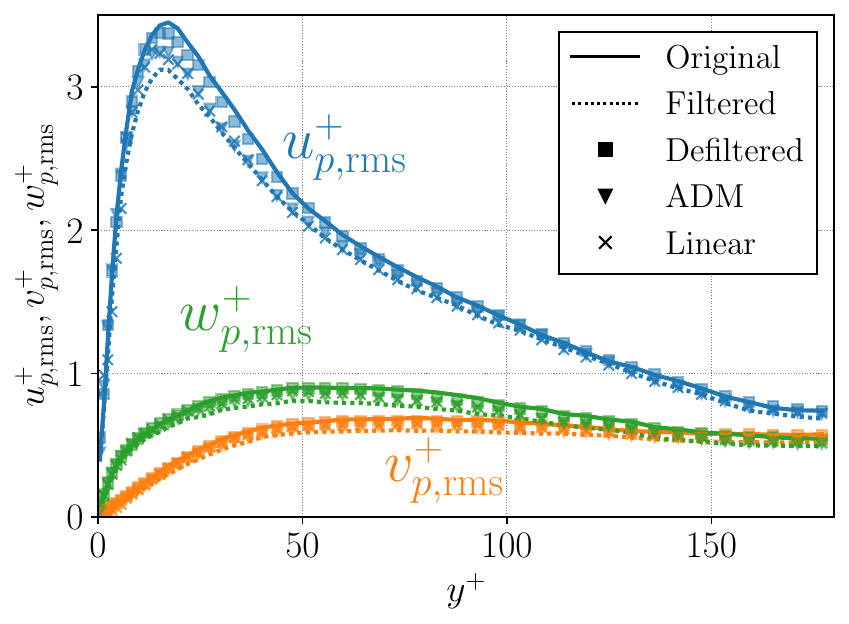}
    \caption{RMS of particle velocity fluctuations in a minimal domain.}
    \label{fig:statistics_of_particle_velocity}
\end{figure}

\begin{figure}[t]
    \centering
    \includegraphics[width=90mm]{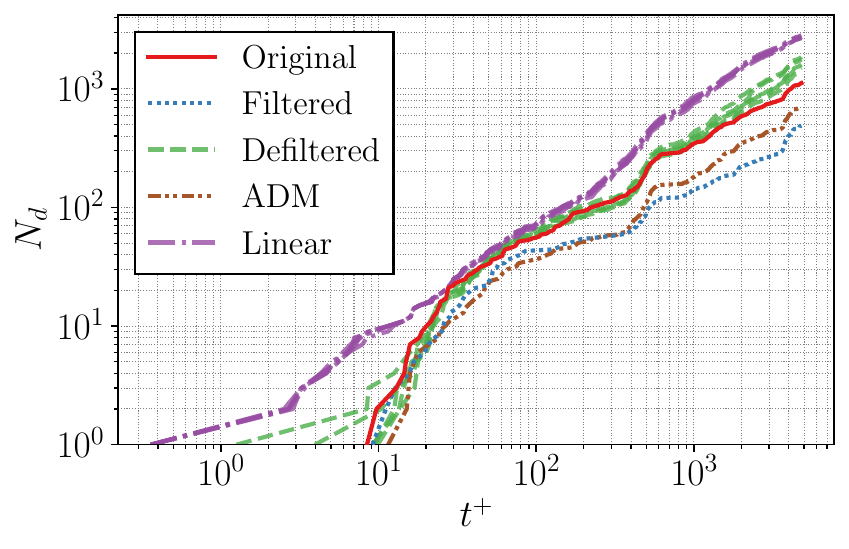}
    \caption{Time evolution of the number of deposited particles, $N_d$. Multiple lines of the defiltered and linear results represent the results of five-fold cross-validation.}
    \label{fig:number_of_deposition}
\end{figure}

Figure~\ref{fig:statistics_of_particle_velocity} shows the RMS of particle velocity fluctuations. As expected, the present method can also recover the particle velocity fluctuations, whereas the accuracy of the linear model and the ADM is insufficient. This result justifies the use of a nonlinear activation function compared with the conventional methods such as the linear optimal schemes~\citep{yeung_algorithm_1988}.

\begin{table}[b]
    \caption{The deposition velocity in wall units, $V^+_d$. 
    The values after $+$ and $-$ signs of the defiltered and the linear results represent a range resulting from five-fold cross-validation.}
    \label{table:deposition_velocity}
    \hspace{100truemm}
    \centering
    {\tabcolsep=5mm
        \begin{tabular}{cc}
            \hline \hline
            Method & Deposition velocity \\ \hline
            Original & $0.040$ \\
            Filtered & $0.019$ \\
            Defiltered & $0.042^{+ 0.004}_{- 0.003}$ \\
            ADM & $0.022$ \\
            Linear & $0.076^{+0.004}_{-0.004}$ \\
            Wood~\citep{wood_mass-transfer_1981} & 0.045\\
            \hline \hline
        \end{tabular}
    }
\end{table}

We subsequently evaluate the number of particles deposited to the channel walls, $N_d$, in a given time period, $t^+=0$--$t^+_d$. Figure~\ref{fig:number_of_deposition} shows $N_d$ in the four cases: original (i.e., fine grid), filtered, defiltered, ADM, and linear (i.e., defiltered using the linear network). Each of the defiltered and linear lines represents the result of five-fold cross-validation. The number of deposited particles in the coarse-grid fields is significantly small, and the result of the ADM is still underestimated. The linear result conversely overestimates the deposition. In contrast, the deposition in the defiltred fields \text{using the present method} shows good agreement with the original one. 

For the quantitative analysis, the deposition velocity in wall units, $V^+_d$, has been commonly used.\citep{li_computer_1993} It is defined as
\begin{equation}
    V^+_{d} = \frac{N_{d} / t^+_{d}}{N / H^+}
    \label{eq:deposition_velocity}
    ,
\end{equation}
where $H^+$ is the channel width in wall units, and $N$ represents the total number of particles introduced at $t^+=0$. Table~\ref{table:deposition_velocity} presents the deposition velocity computed after the flow reached a statistically steady state, i.e., in $t^+ = 100$--$1000$. For reference, an empirical value calculated with the well-known Wood's formula~\citep{wood_mass-transfer_1981} is also presented. As can be seen in the table, the present defiltering method can substantially improve the prediction accuracy of the deposition velocity.

In sum, the present ML-based defiltering method provides great performance to reconstruct the particle characteristics such as the trajectory of an individual particle, the velocity fluctuations, and the deposition velocity.

\begin{figure}[t]
    \centering
    \includegraphics[width=80mm]{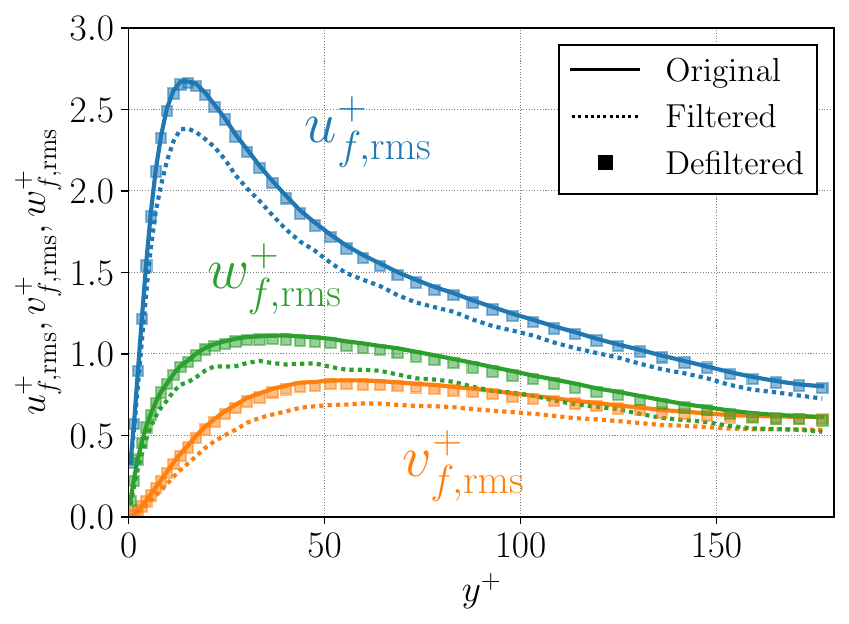}
    \caption{RMS of fluid velocity fluctuations in a larger domain.}
    \label{fig:stat_of_fluid_vel_in_larger_domain}
\end{figure}

\subsection{Statistics of fluid velocity in the larger domain}
\label{sec:fluid_velocity_statistics_in_larger_domain}

Finally, we examine the generalizability of our model, that is, whether the model trained in the minimal domain can be applied to a larger domain (see, Table~\ref{table:computational_domains}). Here, we present only the statistics of fluid velocity because we have already presented in Section~\ref{sec:characteristics_of_tracked_particles} the accuracy of the particle motion when the fluid fields are precisely reconstructed.

The RMS of fluid velocity fluctuations is presented in Fig.~\ref{fig:stat_of_fluid_vel_in_larger_domain}. The defiltered profiles show great agreement with the original ones. This fact implies that the ML model can learn the local characteristics of turbulence precisely enough to reconstruct the interpolated fluid velocity at an arbitrary point. The present results suggests that the training of the ML model can be done at the minimal computational cost using the field in the minimal domain, while the traing ML model can be used in a larger domain ensuring the decay of two-point correlations.

\section{Conclusions}
\label{sec:conclusions}
We proposed an ML-based method for Lagrangian particle tracking in coarse-grid turbulent flow fields. In this method, we use the MLP to reconstruct the fluid velocity at the particle location in coarse-grid fields. First of all, the ML model is trained in the minimal domain of turbulent channel flow so that it can reconstruct the interpolated fluid velocity at a given point accurately. Next, we assessed the present model using the fields which are completely time-uncorrelated but have the same domain size as that used for training the ML model. These results demonstrate the great potential of this model to recover the statistics and characteristics of fluid fields and particles. To be specific, the good agreement in the RMS of fluid and particle velocity fluctuations, the trajectories, and the deposition of particles are presented. Finally, we investigated the generalizability of the present method using a computational domain larger than that used for training the ML model. This test shows the great agreement of the RMS of fluid velocity, which suggests that the training cost of the present ML model can be kept minimal.

As seen above, the present results demonstrate the possibility of ML techniques for precise particle tracking in coarse-grid fields. However, further investigation is required for its use in practical computations. First, the present model does not consider temporal factors as the input. Temporal factors may enable a more accurate reconstruction of turbulent flows especially when a larger time step is used. Second, the {\it a posteriori} test with the LES, aside from the filtered DNS, is required. That is because the LES involves different kinds of errors, such as the subgrid-scale modeling errors, and the accumutation of such errors due to the time integration. Additionally, the way to improve the particle motion in Reynolds-averaged Navier-Stokes (RANS) simulations using ML techniques is also desired, because LPT in RANS simulations is widely conducted for practical applications.

\section*{Acknowledgments}
This work was supported through JSPS KAKENHI (Grant No.~21H05007) by Japan Society for the Promotion of Science.
	
\appendix
\section{Influence of stencil size}
\label{sec:stencil_size_dependence}

\begin{table}[h]
	\caption{The MSE of fluid velocity for each stencil size. The values after $+$ and $-$ signs represent a range resulting from five-fold cross-validation.}
	\label{table:stencil_size}
	\hspace{100truemm}
	\centering
	{\tabcolsep=5mm
		\begin{tabular}{cc}
			\hline \hline
			Stencil size of $(x, y, z)$ & MSE \\ \hline
			$(2, 2, 2)$ & $0.03607^{+0.00006}_{-0.00007}$ \\
			$(3, 3, 3)$ & $0.02675^{+0.00008}_{-0.00007}$ \\
			$(4, 4, 4)$ & $0.02124^{+0.00005}_{-0.00006}$ \\
			\hline \hline
		\end{tabular}
	}
\end{table}

As shown in Section~\ref{sec:model_structure}, the stencil size of ML input, ${\bm q}_{\rm in}$, is set to $(2, 2, 2)$ for the $(x, y, z)$ direction. Here, the stencil size dependence of the reconstruction accuracy is investigated. Table~\ref{table:stencil_size} shows the MSE between the original fluid velocity, ${\bm u}_{f, {\rm fine}}$, and the defiltered fluid velocity, $\widetilde{{\bm u}}_f$ among three different stencil sizes, $(2, 2, 2)$, $(3, 3, 3)$, and $(4, 4, 4)$. This result suggests that a larger stencil can improve the reconstruction accuracy. We emphasize that, in terms of the RMS of ﬂuid velocity and the particle characteristics shown in this study, there is no significant difference depending on the stencil sizes.

\begin{figure}[b]
	\centering
	\includegraphics[width=90mm]{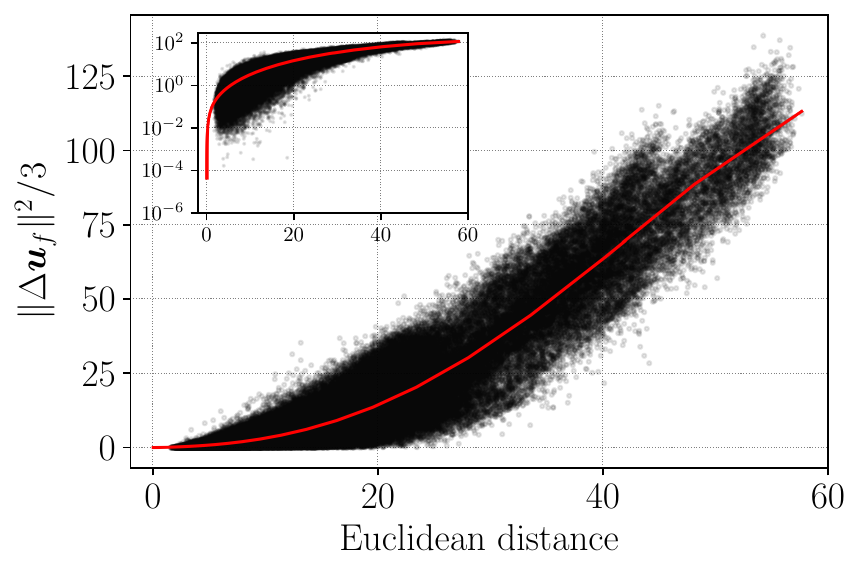}
	\caption{The Euclidean distance of stencil values and the difference in the corresponding fine-grid values. The main figure is on a linear scale while the inset is shown as a semi-logarithmic plot. The red line shows the extrapolation curve with the fifth-degree polynomial.}
	\label{fig:stochastic_uncertainty}
\end{figure}

Next, we evaluate the uncertainty caused by stencils in coarse-grid fields. Although two stencils with identical values (i.e., velocities and pressure) can have different subgrid fluid velocities in the original fields, ${\bm u}_{f, {\rm fine}}$, due to the nature of turbulence, the function $\mathcal{F}_{\rm MLP} \left( {\bm q}_{\rm in} \right)$ always returns the same result with the same stencil values, ${\bm q}_{\rm in}$, in the present method. To evaluate such a stochastic uncertainty, the Euclidean distance between ${\bm q}_{\rm in}$ around the two different points, $\| \bm{q}_{{\rm in}, i} - \bm{q}_{{\rm in}, j} \|$, is calculated. Note that two points are each randomly selected from time-uncorrelated datasets. Figure~\ref{fig:stochastic_uncertainty} shows the Euclidean distance of stencil values and the difference in the corresponding fine-grid values, $\|\Delta \bm{u}_{f, {\rm fine}}\|^2/3 = \left( \bm{u}_{f, {\rm fine}, i} - \bm{u}_{f, {\rm fine}, j} \right)^2 / 3$. Using the polynomial extrapolation, when the Euclidean distance is equal to $0$, i.e., two stencil values are identical, the difference in the corresponding fine-grid values is estimated to be approximately $\mathcal{O}(10^{-6}$--$10^{-5})$, although it depends on the degree of the fitting polynomial. In contrast, the MSE between original and defiltered velocities is $\mathcal{O}(10^{-2})$ in the present method (see, Table~\ref{table:stencil_size}). Thus, the stochastic uncertainty can be ignored in this study.

\section*{Data Availability Statement}
The data that support the findings of this study are available from the corresponding author upon reasonable request.


\bibliographystyle{unsrtnat}
\bibliography{references}  






\end{document}